\documentclass[twocolumn,showpacs,epsfig,pra]{revtex4}
\usepackage{graphicx}
\usepackage{amssymb}
\usepackage{amsmath}

\usepackage{color,soul}

\begin{document}

\draft
\title{Szilard engine revisited; information from time forward and backward process}
\author{Kang-Hwan Kim$^1$ and Sang Wook Kim$^{2}$}
\email{swkim0412@pusan.ac.kr}
\affiliation{$^1$Department of Physics, Pusan National University, Busan 609-735, Korea}
\affiliation{$^2$Department of Physics Education, Pusan National University, Busan 609-735, Korea}
\date{\today}

\begin{abstract}
We derive the work performed in the Szilard engine (SZE) by using dissipative work formula of non-equilibrium thermodynamics developed in Kawai {\it et al.} Phys. Rev. Lett. {\bf 98}, 080602 (2007). The work is described as the difference of probability distributions of measurement outcomes of the time forward and the backward process.

\end{abstract}
\pacs{03.65.Ta,05.70.Ln,89.70.Cf,05.70.-a}

\maketitle
\narrowtext

\begin{figure}
 \vspace*{1.8cm}
 \hspace*{-1.7cm}
  \includegraphics[width=12cm]{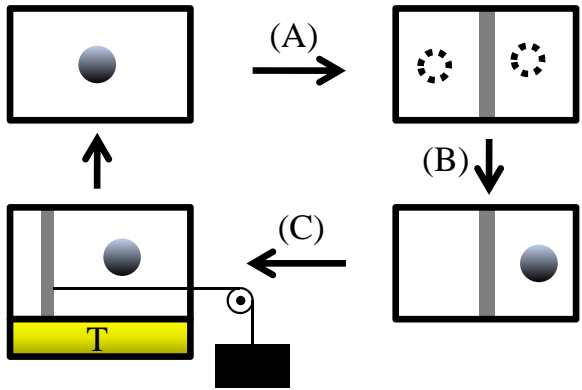}
 \vspace{-5.5cm}
  \caption{Schematic diagram of the thermodynamic processes of the SZE. Initially a single atom is prepared in an isolated box. (A) A wall denoted as a vertical grey bar is inserted to split the box into to two parts. The atom is represented by the dotted circles to reflect that we are lack of the information on which side it is before the measurement. (B) The measurement is performed, and one acquires the knowledge of where the atom is. (C) A load denoted by a filled rectangle is attached to the wall to extract a work via an isothermal expansion at temperature $T$. \label{fig1}}
\end{figure}

Szilard engine (SZE) is machinery that extracts mechanical work from information \cite{Szilard29}. It is made up of an atom (or a molecule) contained in an isolated box. The thermodynamic cycle of the SZE consists of three steps as shown in Fig.~\ref{fig1}; (A) to insert a wall so as to divide a box into two parts, (B) to perform measurement to obtain information on which side the atom is in, and (C) to attach a weight to the wall to extract work via isothermal expansion with a thermal reservoir of temperature $T$ contacted. As the gas is expanded in a quasi-static way, the amount of extracted work is given as $k_BT \ln 2$, where $k_B$ is a Boltzmann constant.

The SZE was originally proposed to show the importance of information in the context of Maxwell's demon \cite{Leff03,Maruyama09}. To avoid violating the second law of thermodynamics the entropy associated with (Shanon) information obtained via measurement process should play an equivalent role as physical (or Boltzmann) entropy \cite{Szilard29,Brillouin51}. In fact, the measurement is assumed to be performed by the Maxwell's demon. It had been shown that the demon can perform measurement reversibly since it can be modeled as one-bit memory in the SZE. The entropy of the engine is then transferred to the demon via measurement. As Landauer pointed out, erasure of the demon's memory should be followed in order to complete the thermodynamic cycle of the engine \cite{Landauer61,Bennett82}. Here the entropy is transferred from the demon to the environment, which is irreversible since the degree of freedom of the environment is large enough. The SZE has been revived in various contexts \cite{Scully03,Kim05,Raizen05}, and realized in experiments \cite{Serreli07,Thorn08,Price08}. The non-equilibrium SZE has also been considered \cite{Sagawa10,Toyobe10}.

The work done by the SZE is directly associated with quantum measurement process, in which information entropy of $\ln 2$ is produced. Conventionally it has been believed that this information forms a source of the SZE. If one considers the SZE consisting of more than one particle, however, such a simple argument needs modification. In this paper we will show that more general statement on the source of work of the SZE can be made by using recent idea developed in non-equilibrium thermodynamics: the work of the SZE is given as the relative-entropy-like formula of measurement outcomes of the time forward and the backward process. It clearly shows that the work formula of the {\em quantum} SZE derived recently \cite{Kim11} has {\em no} quantum mechanical origin since it appears in a purely classical consideration.

The pioneering work on non-equilibrium thermodynamics has been done by Jarzynski \cite{Jarzynski97}. Crooks then found his seminal fluctuation theorem, $p_f (+w) / p_b (-w) = e^{\beta(w-\Delta F)}$, where $p_{f(b)}$ is the probability density function of work in the time forward (backward) direction, and derived Jarzynski equality from it \cite{Crooks98}. This shows the work done during thermodynamic process is intimately related to the time forward and the backward protocol. Recently Kawai, Van den Broeck and Parrondo have found an expression of the average dissipated work $\left< W \right>_{\rm diss}$ upon bringing a system from one canonical equilibrium state described by an external parameter $\lambda_A$ at a temperature $T$ into another one described by $\lambda_B$ at the same temperature \cite{Kawai07}. The dissipated work is defined as the extra amount of work done by a system, on top of the difference of free energy $\Delta F$ required for making this transition. It reads
\begin{equation}
\left< W \right>_{\rm diss}=\left< W \right> + \Delta F = -k_BTD(\rho\parallel\tilde{\rho}),
\label{eq:dissipative work}
\end{equation}
where
\begin{equation}
D(\rho\parallel\tilde{\rho})=\int d\Gamma \rho(\Gamma,t) \ln \frac{\rho(\Gamma,t)}{\tilde{\rho}(\tilde{\Gamma},t)}.
\label{eq:relative entropy}
\end{equation}
Here $\rho = \rho(\Gamma,t)$ is the probability density in phase space to observe the system to be in a micro-state $\Gamma=(q, p)$ specified by a set of positions $q$ and momenta $p$ at an intermediate time $t$. The other density $\tilde{\rho}=\tilde{\rho}(\tilde{\Gamma};t)$ represents the distribution in the time-reversed process observed at a corresponding time-reversed phase point $\tilde{\Gamma}=(q,-p)$ at $t$ measured in the forward process \cite{Kawai07}. $D(\rho\parallel\tilde{\rho})$ in Eq.~(\ref{eq:relative entropy}) is called as the relative entropy quantifying the difference between two distributions $\rho$ and $\tilde{\rho}$ \cite{Vedral02} .

If there exists some constraints on the initial state, which is imposed by selection or filtering processes; for example, an available phase space is restricted to $\Gamma_A$ over $\Gamma$, Eq.(\ref{eq:dissipative work}) is modified as
\begin{equation}
\left< W \right> + \Delta F = - k_B T \ln \frac{p({\Gamma_A})}{\tilde{p}(\tilde{\Gamma}_B)} - k_B T D(\rho\parallel\tilde{\rho}),
\label{eq:dissipative work with feedback}
\end{equation}
where $p({\Gamma_A})$ denotes the probability to select the initial condition within a phase space volume $\Gamma_A$ for the forward process, i.e. $p({\Gamma_A})=Z(T,\lambda_A;\Gamma_A)/Z(T,\lambda_A)$ in equilibrium, while $\tilde{p}(\tilde{\Gamma}_B) = Z(T,\lambda_B;\tilde{\Gamma}_B)/Z(T,\lambda_B)$ for the backward \cite{Parrondo09}. Here the partition function is defined as  $Z(T,\lambda,D) = \int_D d\Gamma \exp \left[ -H(\Gamma,\lambda)/k_BT\right]$, where $H(\Gamma,\lambda)$ is a Hamiltonian.

Although the original SZE consists of only one atom, we will consider $N$-particle SZE since our interpretation becomes more transparent in many particle case. The thermodynamic processes of the SZE are accordingly modified. At first, we obtain $2^N$ outcomes rather than $2$ from the measurement. Note that the particles are distinguishable as classical mechanics is considered. For simplicity we assume that all the measurement is performed perfectly. How the imperfect measurement modifies the results of the SZE has been investigated in Ref. \cite{Sagawa09}. The probability to find $r$ particles in the left side after a wall is inserted at the center of the one-dimensional box of size $L$ is given as
\begin{equation}
p(N,r)=2^{-N} \begin{pmatrix} N  \\ r \end{pmatrix},
\label{eq:p}
\end{equation}
which is related to $p({\Gamma_A})$ in Eq.~(3). The wall then moves to undergo isothermal expansion so that the engine does work until it reaches $x=rL/N$, which is equilibrium position. Finally the wall is removed to return to the initial state. In one-particle SZE, it is not necessary to remove the wall since the wall reaches the end of the box.

\begin{figure}
 \vspace*{2.3cm}
 \hspace*{-1.2cm}
  \includegraphics[width=11cm]{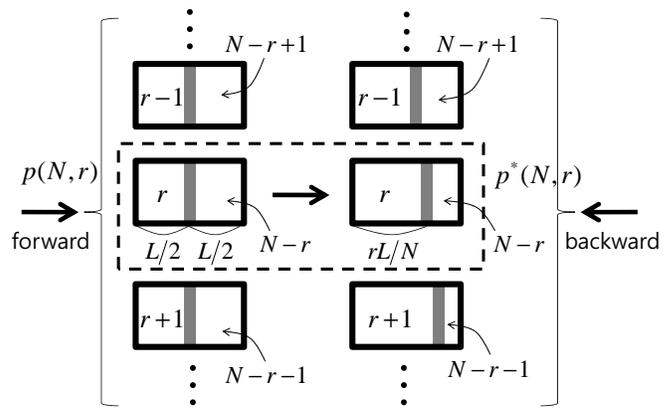}
 \vspace{-5cm}
  \caption{Schematic diagram of the central part of thermodynamic processes of $N$-particle SZE. After inserting a wall at $x=L/2$, $N+1$ possible cases, namely $r=0,1, \cdots N$, are obtained by the measurement. $p(N,r)$ is determined by selecting the case encircled by the dashed box among $N+1$ possible ones. The isothermal expansion then brings the wall to $x=rL/N$. For the backward process, $p^*(N,r)$ is determined by selecting the case encircled by the dashed box, where the wall is located at $x=rL/N$. \label{fig2}}
\end{figure}

For the time reversed process, we start from $N$ particles in a box without the wall, which is the exactly same as the initial condition of the forward process. The wall is then inserted at $x=rL/N$. At this moment the number of particle in the left side can be any number ranging from $0$ to $N$. We should select the $r$ particle case since this is only correct time-reversed process as shown in Fig.~\ref{fig2}. The probability to select such a specific case is given as
\begin{equation}
p^*(N,r)=\left( \frac{r}{N} \right)^r \left( \frac{N-r}{N} \right)^{N-r} \begin{pmatrix} N  \\ r \end{pmatrix}.
\label{eq:p*}
\end{equation}

Now we derive the work performed by the SZE based upon Eq.~(\ref{eq:dissipative work with feedback}).
In order to exploit Eq.~(\ref{eq:dissipative work with feedback}), the thermodynamic process from the initial to the final state should be undergone with the system isolated from the environment, where two states should have the same temperature \cite{Parrondo09}. Since the expansion process is performed isothermally, the system is not isolated but rather equilibrated with the reservoir at every moment. However, the {\em quasi-static} isothermal expansion process performed during the time interval $[0,\tau_a]$ can be regarded as a collection of many consecutive {\em isolated} subprocesses, namely $[0,\epsilon], [\epsilon,2\epsilon], \cdots, [(n-1)\epsilon,n\epsilon], \cdots,$ and $[\tau_a-\epsilon,\tau_a]$ ($\epsilon \ll \tau_a$) with $n=1,2, \cdots, M$. Here at $t=0$ and at $t=\tau_a$ the expansion process starts and ends, respectively. It is emphasized that at the end of each subprocess the system should contact the reservoir of temperature $T$ to ensure that the isothermal condition is guaranteed. This is also necessary for exploiting Eq.~(\ref{eq:dissipative work with feedback}) \cite{Parrondo09}. We thus assume that Eq. (\ref{eq:dissipative work with feedback}) is satisfied within each individual subprocess:
\begin{equation}
\left< W_n \right> + \Delta F_n = - k_B T \ln \frac{p({\Gamma_{A,n}})}{\tilde{p}(\tilde{\Gamma}_{B,n})} - k_B T D(\rho_n\parallel\tilde{\rho_n}).
\label{eq:work_n}
\end{equation}

When the expansion is completed, the wall is removed at $t = \tau_b (>\tau_a)$. It is not so crucial how fast the wall is removed in classical consideration, which is not true in quantum mechanics \cite{Kim11}. We again need a relaxation process by contacting a reservoir of temperature $T$ at a certain time $t=\tau_c (> \tau_b)$. The reason is that although the wall is removed the distribution $\rho (=\rho_N)$ cannot cover whole accessible phase space given by the canonical distribution of temperature $T$ due to conservation of the phase volume followed by Liouville dynamics. Such an additional relaxation at the end of the cycle plays a crucial role in our consideration. This will be discussed in detail below.

Now we consider each term appearing in Eq. (\ref{eq:dissipative work with feedback}) one by one. At first, the second term $\Delta F$ should vanish for one full cycle since the SZE is a cyclic engine, implying that the initial state coincides with the final state. Secondly, one can show that $p({\Gamma_{A,n+1}})=\tilde{p}({\tilde{\Gamma}_{B,n}})$ is obeyed during $[0,\tau_a]$ from the fact that at the end of each subprocess, namely $t=n\epsilon$, the system is equilibrated with the reservoir of temperature $T$. It leads us to $\sum_{n=1}^{M} \ln \left[ p({\Gamma_{A,n}}) / \tilde{p}(\tilde{\Gamma}_{B,n}) \right] = \ln p({\Gamma_{A,1}}) - \ln \tilde{p}(\tilde{\Gamma}_{B,N})$. On the other hand, during $[\tau_a,\tau_c]$ one finds $p({\Gamma_{A,ac}})=\tilde{p}({\tilde{\Gamma}_{B,N}})$ and $\tilde{p}({\tilde{\Gamma}_{B,ac}})=1$, where the subscript $ac$ denotes a subprocess during $[\tau_a,\tau_c]$. Thus one has merely $-k_B T \ln p({\Gamma_{A,1}})$ or equivalently $-k_B T \ln p(N,r)$ for the thrid term of Eq. (\ref{eq:dissipative work with feedback}). Lastly, during $[0,\tau_a]$ it is shown that $\sum_{n=1}^M D(\rho_n\parallel\tilde{\rho_n})$ vanishes as $M \rightarrow \infty$. The reason is that as $M \rightarrow \infty$ the whole expansion process approaches true isothermal process, where $\rho_n = \tilde{\rho_n}$ is satisfied. More rigorously one can prove  $\sum_{n=1}^M D(\rho_n\parallel\tilde{\rho_n}) \sim 1/M$. On the other hand, during $[\tau_a,\tau_c]$ one finds $\rho_{ac} / \tilde{\rho}_{ac} = 1/p^*(N,r)$ since (i) $\rho_{ac}$ is equal to $\rho_N$, whose phase volume is confined in $\Gamma_A$ determined from the measurement outcome $r$, and (ii) $\tilde{\rho}_{ac}$ is the canonical distribution of temperature $T$ at $t=\tau_c$, which means it covers whole available phase volume $\Gamma$. Thus the fourth term of Eq. (\ref{eq:dissipative work with feedback}) becomes simply $+k_B T \ln p^*(N,r)$ due to $D(\rho_{ac} || \tilde{\rho}_{ac}) = \int d\Gamma \rho_{ac} \ln \left[ 1/p^*(N,r) \right]$. In fact, such an entropy production described by $D(\rho_{ac} || \tilde{\rho}_{ac})$ comes from the fact that the wall removal induces an {\em irreversible} process, namely free expansion.

So far we have derived the work performed for the case that one specific selection, namely $\Gamma_A$ or a measurement outcome $r$ in our case, is made. The {\em average} work for all possible outcomes is then expressed as
\begin{equation}
\left< W \right> = -k_B T \sum_{r=0}^N p(N,r) \ln \frac{p(N,r)}{p^*(N,r)}.
\label{eq:work}
\end{equation}
This looks like the relative entropy, but indeed it is not the case since $p^*(N,r)$ is not normalized, namely $\sum_{r=0}^N p^*(N,r) \neq 1$. Note that this is distinguished from the following normalization condition.
\begin{equation}
\sum_{m=0}^N \left( \frac{r}{N} \right)^m \left( \frac{N-r}{N} \right)^{N-m} \begin{pmatrix} N  \\ m \end{pmatrix}=1.
\end{equation}
In Eq.~(\ref{eq:work}) $p(N,r)$ is the selection probability of the forward measurement, while $p^*(N,r)$ corresponds to that of the backward if the measurement is assumed to be performed for the backward. It should be noted that originally $p^*(N,r)$ comes from the entropy production induced by removing the wall, i.e. $D(\rho_{ac} || \tilde{\rho}_{ac})$.

In a trivial one-particle SZE, the well-known work, $k_B T \ln 2$, is retrieved from two probability distributions $p(1,0)=p(1,1)=1/2$, and $p^*(1,0)=p^*(1,1)=1$ by using Eq.~(\ref{eq:work}) since for the backward process the wall is inserted at $x=0$ ($x=L$), only $r=0$ ($r=1$) can be selected. Conventionally the SZE performs work by exploiting the entropy produced during the measurement of the forward process, where the backward is ignored. In one particle SZE, however, the contribution of the backward process is invisible.

In two-particle SZE, the work is also given as $k_B T \ln 2$ from $p(2,0)=p(2,2)=1/4$, $p(2,1)=1/2$, $p^*(2,0)=p^*(2,2)=1$, and $p^*(2,1)=1/2$. In a conventional point of view, the information entropy produced by the measurement is $\sum_r p(2,r) \ln p(2,r) = 2 \ln 2$, so that naively thinking it is expected that $\left< W \right> = 2k_B T \ln 2$. However, in the case of $r=1$ work cannot be generated because there is no pressure difference between two sides. According to $p(2,1)=1/2$ the average work is thus given as $k_B T \ln 2$ rather than $2 k_B T \ln 2$. Since all the information entropy related to the forward measurement is not used for generating work, it is necessary to eliminate useless information for work. In fact, the selection probability of the backward process exactly plays a role of useless information. In this sense the SZE is machinery that extracts work from the difference of information between the forward and the backward process.

By using Eqs.~(\ref{eq:p}) and (\ref{eq:p*}), Eq.~(\ref{eq:work}) is rewritten as
\begin{equation}
\left< W \right> = \sum_r p(N,r) w(N,r)
\end{equation}
with
\begin{equation}
w(N,r) = N k_B T \left[ \ln 2 - b(r/N)  \right],
\label{eq:work_info}
\end{equation}
where $b(q)= - q \ln q - (1-q) \ln (1-q)$ is a binary entropy function satisfying $0 \le b(q) \le \ln 2$. The entropy of the system decreases by $\ln 2$ due to the measurement, while it increases by $b(q)$ when the wall is removed. The former is the information entropy, but the latter is the physical entropy divided by the Boltzmann constant, acquired by removing a partition. Net entropy change of the system is then given as the difference between these two, namely $\Delta S$. The maximum work that the SZE can generate is then at best $k_B T \Delta S$, i.e. Eq.~(\ref{eq:work_info}).

Eq.~(\ref{eq:work_info}) can also be retrieved from classical thermodynamic consideration as follows
\begin{equation}
\frac{w(N,r)}{k_B T} = \int^{qL}_{L/2} \frac{r}{V} dV + \int^{(1-q)L}_{L/2} \frac{(N-r)}{V} dV ,
\label{eq:work_thermo}
\end{equation}
where $q$ denotes $r/N$. Here the first and the second term on the right-hand side represent the work done by the gases in the left and the right side, respectively, during the isothermal expansion. However, from Eq.~(\ref{eq:work_thermo}) it may not be easy to draw information-theoretic interpretation presented in Eq.~(\ref{eq:work}), i.e. the difference of information between the forward and the backward process. More importantly Eq.~(\ref{eq:work}) is still valid in the quantum version of the SZE, while Eq.~(\ref{eq:work_thermo}) is not any longer \cite{Kim11}.

As $N \rightarrow \infty$ the work per a particle vanishes since it is approximated for $N \gg 1$ as
\begin{equation}
\frac{\left< W \right>}{N} = k_B T \left( \frac{1}{2} \frac{1}{N} + \frac{1}{12} \frac{1}{N^2} + \frac{1}{30} \frac{1}{N^3} + \cdots \right).
\end{equation}
This can be understood from the fact that it is almost improbable to have considerable number-difference of particles between two sides compared with $N( \gg 1)$ after inserting a wall. As a matter of fact, the maximum work is obtained from one-particle SZE.

So far, for simplicity we have considered the wall is inserted at $x=L/2$ in the forward process. If we take $x=d$ $(0 \le d \le L)$ into account, Eqs.(\ref{eq:p}) and (\ref{eq:work_info}) are replaced by
\begin{equation}
p(N,r)=\left( \frac{d}{L} \right)^r \left( \frac{L-d}{L} \right)^{N-r} \begin{pmatrix} N  \\ r \end{pmatrix},
\end{equation}
and $w(N,r) = N k_B T \left[ b(d/L) - b(r/N)  \right]$, respectively. Thus the work is given as difference of two entropy functions of the forward and the backward process.

Recently the formula of work performed by the {\em quantum} SZE is found based upon quantum thermodynamic approach \cite{Kim11}. This expression exactly coincides with Eq.~(\ref{eq:work}). In some sense this may not be surprising since Eq.~(\ref{eq:dissipative work}) is proven to be valid in quantum mechanics \cite{Parrondo09}. However, this is not trivial because it was found that the work is required for inserting or removing a wall in quantum mechanics \cite{Kim11} even though no work is needed in classical case. The difference between the classical and the quantum SZE indeed lies at their partition functions from which the probabilities $p$ and $p^*$ are determined both in classical and in quantum mechanics, i.e.
$p(N,r) = {Z_N(r,d/L;T)} / {\sum_r Z_N(r,d/L;T)}$ and $p^*(N,r) = {Z_N(r,rL/N;T)} / {\sum_r Z_N(r,rL/N;T)}$,
where $Z_N(r,x;T)$ denotes a $N$-particle partition function describing the case that $r$ particles of temperature $T$ exist in the left of the wall located at $x$. Different from classical mechanics, particles are identical in quantum mechanics, which dramatically affects the partition functions. It has been found that more (less) work can be extracted from the bosonic (fermionic) SZE compared with that of the classical SZE. Irrespective of such a marked difference the work formula has the equivalent form, i.e. Eq.~(\ref{eq:work}), and the same physical interpretation. 

In conclusion we have shown that the work performed by many-particle SZE is given as the relative-entropy-like formula describing the difference of the probability distributions of the time forward and the backward process. This result comes from the dissipative work formula of non-equilibrium thermodynamics. The expression obtained here is also valid in quantum mechanics. We believe our finings shed light on the subtle role of information in physics.

We would like to thank Juan Parrondo, Takahiro Sagawa, Simone De Liberato, and Masahito Ueda for useful discussions. This was supported by the NRF grant funded by the Korea government (MEST) (No.2009-0084606, No.2009-0087261 and No.2010-0024644).

\end{document}